\documentstyle[preprint,aps,prb]{revtex}

\draft

\begin{document}
\title{Functions of linear operators: Parameter differentiation}
\author{Domingo Prato$^{1,2}$ and Constantino Tsallis $^2$}
\address{$^1$Facultad de Matematica, Astronomia y Fisica, Universidad Nacional de\\
Cordoba, Ciudad Universitaria, 5000 Cordoba, Argentina\\
$^2$Centro Brasileiro de Pesquisas Fisicas (CBPF),\\
Rua Dr. Xavier Sigaud 150, 22290-180 Rio de Janeiro, Brazil\\
prato@mail.famaf.unc.edu.ar \\
tsallis@cbpf.br}
\maketitle

\begin{abstract}
We derive a useful expression for the matrix elements $\bigl[\frac{\partial
f[A(t)]}{\partial t}\bigr]_{i\;j}$ of the derivative of a function $f[A(t)]$
of a diagonalizable linear operator $A(t)$ with respect to the parameter $t$.
The function $f[A(t)]$ is supposed to be an operator acting on the same
space as the operator $A(t)$. We use the basis which diagonalizes $A(t)$, i.e., 
$A_{i\;j}=\lambda_i\; \delta_{i\;j}$, and obtain 
$\Bigl[\frac{\partial f[A(t)]}{\partial t}\Bigr] _{i\;j}=\bigl[\frac{\partial A}{\partial t}\bigr]_
{i\;j}\frac{ f(\lambda_j)\;-\;f(\lambda_i)}{\lambda_j\;-\;\lambda_i}$. In addition to
this, we show that further elaboration on the (not necessarily simple)
integral expressions given by Wilcox 1967 (who basically considered $f[A(t)]$
of the exponential type) and generalized by Rajagopal 1998 (who extended
Wilcox results by considering $f[A(t)]$ of the $q$-exponential type where
 $\exp_q(x) \equiv [1+(1-q)x]^{1/(1-q)}$ with $q \in {\cal {R}}$; hence, 
$\exp_1 (x)=\exp(x))$ yields this same expression. Some of the lemmas first
established by the above authors are easily recovered.\\
\end{abstract}


\bigskip

The quite ubiquitous necessity to have an expression of the parameter
derivative of a general function of a linear operator has stimulated many
authors to develop various algorithms focusing on this problem. Special
effort has addressed the calculation of the derivative of exponential
functions of such operator in terms of the derivative of the exponent. Most
of the results that have been achieved provide the derivative as an integral
expression\cite{wilcox,rajagopal}, or by means of an integral representation 
\cite{poincare} of the function in the complex plane. Such elegant procedure
is very useful to prove mathematical properties but, in many cases, to
calculate the remaining integrals is not an easy task. Therefore, it is
convenient to have explicit expressions of the matrices elements of the
relevant operators. The choice of an appropriate basis set considerably
simplifies the calculations; indeed,  once these elements are obtained, the expression
in a different basis can readily be calculated. This is the purpose of the present
work.

Let $A(t)$ and $A^{\prime }(t)\equiv \partial A(t)/\partial t$ respectively
be a general diagonalizable linear operator and its $t$-parameter derivative
($t$ does not necessarily denote time); in general they do no commute. Also,
let $f[A(t)]$ be an arbitrary, well behaved, function of the operator. We want 
to calculate $\frac{\partial f[A(t)]}{\partial t}$. For convenience, we shall 
express the matrices of all operators in the basis which diagonalizes $A(t)$. 
More specifically, we have that 

\begin{equation}
\lbrack A(t)]_{i\;j}=\lambda _{i}\;\delta _{i\;j}
\end{equation}
where the $\{\lambda _{i}\}$ are the eigenvalues of $A(t)$. For instance, in the usual 
case of perturbation theory in Quantum Mechanics we have

\begin{equation}
          H(t)=H_{0}+t \; H_{1}
\end{equation}

with $H_{0}\;=\;H(t=0)$ the unperturbed diagonal Hamiltonian, and $H_{1}$ the perturbation 
of the system. Obviously for this case we have

\begin{equation}
\frac{\partial H(t)}{\partial t}=H_{1}
\end{equation}

We shall start by considering a simple case, namely $f[A]=A^{r}$ with $r\in 
{\cal {R}}$. The derivative of the power $r$ of the linear operator $A(t)$ is by
definition

\begin{equation} 
\frac{\partial A^{r}}{\partial t}\equiv \lim_{s \rightarrow 0}
 \frac{[A(t+s)]^r-[A(t)]^r}{s}    \label{derpot}
\end{equation}

In order to find an expression for $[A(t+s)]^r$ it is convenient to find the basis in which
$A(t+s)$ is diagonal. More precisely, we have

\begin{equation}
A(t+s)\;C\;=\;C\;\Lambda
\end{equation}
where $\Lambda$ is the diagonal matrix formed with the eigenvalues of $A(t+s)$
and $C$ is the matrix whose columns are the eigenvectors also of the matrix $A(t+s)$, hence

\begin{eqnarray}
A(t+s) & = & A(t)+A^{\prime}(t)\;s+...=\lambda+A^{\prime}(t)\;s+...\nonumber\\
C & = & I+C^{(1)}\;s+.....\\      \label{diagonal}
\Lambda & = & \lambda+\lambda^{(1)}\;s+..\nonumber
\end{eqnarray}
where we have kept terms up to order $s$,  $I$ is the matrix unity and $\lambda$ denotes 
the diagonalized matrix $A(t)$. Identification of the coefficients of $s$ implies

\begin{equation}
A^{\prime} +\lambda\;C^{(1)}=C^{(1)}\;\lambda+\lambda^{(1)}
\end{equation} 
Then, by taking matrix elements in the last equation, we obtain

\begin{eqnarray}
A^{\prime}_{i\;i} &=&\lambda^{(1)}_{i}\nonumber\\
A^{\prime}_{i\;j}+\lambda_{i}\;C^{(1)}_{i\;j}&=&C^{(1)}_{i\;j}\;\lambda_{j}\;\;\;\;\;(i \neq j) 
\end{eqnarray}
hence

\begin{eqnarray}
C^{(1)}_{i\;j}&=&\frac{A^{\prime}_{i\;j}}{[\lambda_{j}-\lambda_{i}]}\nonumber  \label{eigevect}
\end{eqnarray}
Once we know the expression for $C$ as a function of $s$ we can obtain an expression for
$[A(t+s)]^r$, namely

\begin{eqnarray}
[A(t+s)]^r=C\;\Lambda^{r}\;C^{-1}
\end{eqnarray}
Then, by keeping up to terms linear in $s$, we can write

\begin{equation}
A^{r}(t+s)\sim [I+s\;C^{(1)}]\;[\lambda+s\;\lambda^{(1)}]^{r}\;[I+s\;C^{(1)}]^{-1}
\end{equation}
and, since we have that $[I+s\;C^{(1)}]^{-1}=I-s\;C^{(1)} + \theta(s^{2})$,  we can
replace this in Eq.(\ref{derpot}) and obtain

\begin{eqnarray}
\frac{\partial [A(t)]^r}{\partial t} &=& \lim_{s\rightarrow 0}\frac{[I+s\;C^{(1)}]\;
[\lambda+s\;\lambda^{(1)}]^{r}\;[I-s\;C^{(1)}]-\lambda^{r}}{s}\nonumber\\\\
&=& C^{(1)}\;\lambda^{r}-\lambda^{r}\;C^{(1)}+r\;\lambda^{(r-1)}\;\lambda^{(1)}\nonumber
\end{eqnarray}
The matrix elements obtained from the last equation are

\begin{eqnarray}
\Bigl[ \frac{\partial [A(t)]^r}{\partial t}\Bigr]_{i\;i} &=&  r\;(\lambda_{i})^{(r-1)}\;\lambda^{(1)}_{i}
=r\;(\lambda_{i})^{(r-1)}\;A^{\prime}_{i\;i}\nonumber\\\\
\Bigl[\frac{\partial [A(t)]^r}{\partial t}\Bigr]_{i\;j} &=&  C^{(1)}_{i\;j}\;
[(\lambda_{j})^{r}-(\lambda_{i})^{r}]=A^{\prime}_{i\;j} \; \; \frac{(\lambda_{j})^{r}-
(\lambda_{i})^{r}}{\lambda_{j}-\lambda_{i}}\;\;\;\;(i \ne j) \nonumber
\end{eqnarray}

where we have used that  $C^{(1)}_{i\;j}\;=\;\frac{A^{\prime}_{i\;j}}{\lambda_{j}-\lambda_{i}}$.

The straightforward application of the above procedure to an arbitrary well behaved function of 
the linear operator $f[A(t)]$ leads us to the following expressions:

\begin{eqnarray}
\Bigl[\frac{\partial f[A(t)]}{\partial t}\Bigr]_{i\;i} =
A^{\prime}_{i\;i} \; \; \left.\frac{\partial f(\lambda)}{\partial \lambda}\right|_{\lambda=
\lambda_{i}}
\nonumber\\\\ 
\Bigl[\frac{\partial f[A(t)]}{\partial t}\Bigr]_{i\;j} = A^{\prime}_{i\;j} \; \; \frac{f(\lambda_{j})-
f(\lambda_{i})}{\lambda_{j}-\lambda_{i}}\;\;\;\;(i \ne j)\nonumber 
\end{eqnarray}

An alternative proof of these same results can be obtained as follows\cite{rajagopal2}. 
Considering that any function $f[A(t)]$ commutes with the operator $ A(t)$, i.e.
$ [\;A(t),\;f(A)\;]=0$, and then taking the derivative with respect to the parameter $t$ of this
commutator, and finally exhibiting the matrix elements, we also arrive to Eqs. (13).
These two equations constitute the main result of this work. In what follows, and in 
order to exhibit their usefulness, we will use these expressions to prove some of the 
theorems given in Wilcox's and Rajagopal's papers.

As a first application let us consider the formula (2.1) of Wilcox \cite{wilcox}, namely

\begin{equation}
\frac{\partial \exp[A(t)]}{\partial t}\;=\;\int_{0}^{1}dx\;\exp[A(1-x)]\; \frac{\partial A(t)}{\partial t}\; \exp(xA).\\
\end{equation}

The matrix elements in the basis that diagonalizes the operator $A(t)$ are given by

\begin{eqnarray}
\Bigl[\frac{\partial \exp[A(t)]}{\partial t}\Bigr]_{i\;j}\;&=&\;\int_{0}^{1}dx\;\exp[(1-x)\lambda_{i}]\;
\Bigl[\frac{\partial A(t)}{\partial t}\Bigr]_{i\;j}\; \exp(x\lambda_{j})\nonumber\\ \nonumber\\
&=&\;\Bigl[\frac{\partial A(t)}{\partial t}\Bigr]_{i\;j}\;\exp(\lambda_{i})\;\int_{0}^{1}\;dx\;\exp[x(\lambda_{j}-\lambda_{i})]\;\nonumber\\\\
&=&\;\Bigl[\frac{\partial A(t)}{\partial t}\Bigr]_{i\;j}\;\frac{\exp(\lambda_{j}) - \exp(\lambda_{i})}{\lambda_{j}-\lambda_{i}}\nonumber\\\nonumber\\
&=& A^{\prime}_{i\;j}\;\frac{\exp(\lambda_{j}) - \exp(\lambda_{i})}{\lambda_{j}-\lambda_{i}},\nonumber
\end{eqnarray}

which is the same expression obtained from our general formula, Eq. (13).

As a second application of our results, let us focus on the proof of the following lemma:

$\;\;\;\;\;\;\;\;\;\;\;\;\;$if   $ [B, A(t)]\; =\; \frac{\partial A}{\partial t}\; 
$  then  $\;[B, f(A)]\;=\; \frac{\partial f(A)}{\partial t}$.\\

Proof: from the first equation we have 

\begin{equation}
B_{i\;j}\;\lambda_{j}\;-\;\lambda_{i}\;B_{i\;j}\;=\; B_{i\;j}(\lambda_{j}\;-\lambda_{i})\;=\;A^{\prime}_{i\;j}
\end{equation}
Taking the matrix elements of the first member of the second equation we get

\begin{equation}
B_{i\;j}\Bigl[f(\lambda_{j})\;-\;f(\lambda_{i})\Bigr]\;=\;\frac{A^{\prime}_{i\;j}}{(\lambda_{j}-\lambda_{i})}\Bigl[f(\lambda_{j})\;-\;f(\lambda_{i})\Bigr]=
\;\Bigl[\frac{\partial f(A)}{\partial t}\Bigr]_{i\;j}
\end{equation}
that is what we wanted to prove.

As a third application, let us now assume that the operator $A$ depends on two parameters, 
namely $t$ and $u$. We will prove the following theorem

\begin{equation}
Tr\Bigl[h(A)\frac{\partial f(A)}{\partial t}\frac{\partial g(A)}{\partial u}\Bigr]=
Tr\Bigl[h(A)\frac{\partial g(A)}{\partial t}\frac{\partial f(A)}{\partial u}\Bigr]
\end{equation}
where $h,f$ and $g$ are arbitrary functions of the operator $A(t,u)$.\\
Proof: the first term of the last equation is
\newpage
\begin{eqnarray}
\sum_{i}\sum_{j}\;h(\lambda_{i})\;\Bigl[\frac{\partial f(A)}{\partial t}\Bigr]_{i\;j}\;\vspace{8 mm}
\Bigl[\frac{\partial g(A)}{\partial u}\Bigr]_{j\;i} \vspace{8 mm} \nonumber\\
=\sum_{i}\sum_{j}\;h(\lambda_{i})\;\frac{A^{\prime}_{i\;j}}{(\lambda_{j}-\lambda_{i})}\;
\Bigl[f(\lambda_{j})\;-\;f(\lambda_{i})\Bigr]\;\frac{\dot{A}_{j\;i}}{(\lambda_{i}-\lambda_{j})}\;
\Bigl[g(\lambda_{i})\;-\;g(\lambda_{j})\Bigr] \nonumber\\\nonumber\\
=\sum_{i}\sum_{j}\;h(\lambda_{i})\;\frac{A^{\prime}_{i\;j}}{(\lambda_{j}-\lambda_{i})}\;
\Bigl[g(\lambda_{j})\;-\;g(\lambda_{i})\Bigr]\;\frac{\dot{A}_{j\;i}}{(\lambda_{i}-\lambda_{j})}\;
\Bigl[f(\lambda_{i})\;-\;f(\lambda_{j})\Bigr]\\
=\sum_{i}\sum_{j}\;h(\lambda_{i})\;\Bigl[\frac{\partial g(A)}{\partial t}\Bigr]_{i\;j}\;
\Bigl[\frac{\partial f(A)}{\partial u}\Bigr]_{j\;i} \nonumber\\ \nonumber\\
=Tr\Bigl[h(A)\;\frac{\partial g(A)}{\partial t}\;\frac{\partial f(A)}{\partial u}\Bigr]\nonumber\\ \nonumber
\end{eqnarray}
as we wanted to prove. (We have used the notation $A^{\prime} \equiv \frac{\partial A}
{\partial t}$ and $\dot{A} \equiv \frac{\partial A}{\partial u}$.)

Let us finally focus on our fourth and last application. As mentioned before, 
Rajagopal\cite{rajagopal} has generalized Wilcox's expression in which the exponential function 
of the operator is replaced by a monomial fractional power of the form

\begin{equation}
Q_{T}(t,\beta)\;=\;\Bigl[1\;-\;(1-q)\beta A(t)\Bigr]^{\frac{q}{1-q}}\;
\equiv\;\Bigl[exp_{q}[-\beta A(t)]\Bigr]^{q}
\end{equation}
Such q-exponential expressions arise naturally when we consider ensembles in the context of 
the recently introduced nonextensive thermostatistics\cite{tsallis} (for a recent review on 
the subject see \cite{review}).

Before we show that further elaboration on Rajagopal's expression\cite{rajagopal} gives the 
same result as the one obtained from our Eq. (13), we need the following identity

\begin{equation}
\frac{d}{dx}\Bigl[\frac{1-ax}{1-bx}\Bigr]^{c}\;=\;c\;(b-a)\frac{[1-ax]\;^{c-1}}{[1-bx]\;^{c+1}} 
\label{identidad}
\end{equation}

Defining again $A^{\prime} \equiv \frac{\partial A(t)}{\partial t} $, Rajagopal's equality, 
besides a factor $\beta$ that is lacking in his equation and also in Wilcox's paper, can be 
written in the form

\begin{eqnarray}
  \vspace{15 mm} \frac{\partial Q_{T}(t,\beta)}{\partial t} \hspace{20 mm} \nonumber\\
=-q\;\beta \int_{0}^{\beta}dxQ_{T}(t,\beta)[Q_{T}(t,x)]^{-1}
 \bigl[1-(1-q)xA(t) \Bigr]^{-1} A^{\prime} \Bigl[1-(1-q)xA(t) \Bigr]^{-1} Q_{T}(t,x)
\end{eqnarray}
The matrix elements of the last equality give us

\begin{eqnarray}
\Bigl[\frac{\partial Q_{T}(t,\beta)}{\partial t}\Bigr]_{i\;j}\;\;\;\;\; \nonumber\\
= -q\;\beta \int_{0}^{\beta}dx\Bigl[1-(1-q) \beta \lambda_{i} \Bigr]^{\frac{q}{1-q}} 
\Bigl[1-(1-q)x \lambda_{i} \Bigr]^{-\frac{q}{1-q}-1} A^{\prime}_{i\;j} 
\Bigl[1-(1-q)x\lambda_{j} \Bigr]^{\frac{q}{1-q}-1}
\end{eqnarray} 
Using the identity given in Eq.(\ref{identidad}) we obtain

\begin{eqnarray}
\Bigl[\frac{\partial Q_{T}(t,\beta)}{\partial t}\Bigr]_{i\;j}\;\;\;\;\; \nonumber\\ \nonumber\\
= -q\;\beta  \Bigl[1-(1-q) \beta \lambda_{i} \Bigr]^{\frac{q}{1-q}} A^{\prime}_{i\;j}
\int_{0}^{\beta}dx \frac{\Bigl[1-(1-q)x\lambda_{j} \Bigr]^{\frac{q}{1-q}-1}}
{\Bigl[1-(1-q)x \lambda_{i} \Bigr]^{\frac{q}{1-q}+1}}\\
= \frac{A^{\prime}_{i\;j}}{\lambda_{j}-\lambda_{i}} \;
\Bigl[[1-(1-q)\beta\lambda_{j} ]^{\frac{q}{1-q}} -
 [1-(1-q)\beta\lambda_{i} ]^{\frac{q}{1-q}} \Bigr] \nonumber\\ \nonumber
\end{eqnarray}
that is the expression that we would obtain from Eq.(13) applied to the 
function, $Q_{T}(t,\beta)$, defined in Eq.(20).

Summarizing, we have given a general expression, Eq.(13), for the parameter 
differentiation of a generic function $f(A)$ of a diagonalizable linear operator $A(t)$ that
depends on a parameter $t$. This expression enables the consistent recovering, as particular instances, of various useful Wilcox formulae\cite{wilcox}, as well  as their recent generalization by Rajagopal\cite{rajagopal}. We believe, consequently, that the present expression can be useful for quantum calculations in standard or deformed Quantum Mechanics, extensive or nonextensive Statistical Mechanics\cite{tsallis}, among other areas of physical interest.

We are deeply indebted to A.K. Rajagopal for a critical reading of our manuscript. This resulted in the beautiful alternative proof presented just after Eqs. (13), which we are including here thanks to his generous authorization.
Partial support from CNPq and PRONEX/FINEP (Brazilian agencies) are
acknowledged as well. One of us (D. P.) also acknowledges warm hospitality at the Centro 
Brasileiro de Pesquisas Fisicas.


\begin{references}

\bibitem{wilcox}  R.M. Wilcox, J. Math. Phys. {\bf 8}, 962 (1967).

\bibitem{rajagopal}  A.K. Rajagopal, Braz. J. Phys. {\bf  29}, 61 (1999);

[http://sbf.if.usp.br/WWW$_{-}$pages/Journals/BJP/Vol29/Num1/index.htm]

\bibitem{poincare}  H. Poincare, Trans. Cambridge Phil. Soc. {\bf 18}, 220 (1899).

\bibitem{rajagopal2} A.K. Rajagopal, private communication (1999).

\bibitem{tsallis}  C. Tsallis, J. Stat. Phys. {\bf 52}, 479 (1988); for a regularly updated 

bibliography see  http://tsallis.cat.cbpf.br/biblio.htm

\bibitem{review}C. Tsallis, Braz. J. Phys. {\bf 29}, 1 (1999); 

[http://sbf.if.usp.br/WWW$_{-}$pages/Journals/BJP/Vol29/Num1/index.htm]

\end{references}
\end{document}